\documentclass[12pt]{article}
\usepackage{ascmac}
\usepackage{ulem}
\usepackage{latexsym}
\usepackage[pdftex]{graphicx,color}
\usepackage{here}
\usepackage{amsmath}
\usepackage[top=3.5cm,bottom=3.5cm,left=2cm,right=2cm]{geometry}
\usepackage{pifont}          
\usepackage{amsmath,amssymb}
\usepackage{mathrsfs}
\usepackage{comment}
\usepackage{geometry}
\usepackage{caption}
\usepackage{ulem}
\newcommand{\paren}[1]{\left(#1\right)}

\newcommand{\mr}{\mathrm}
\newcommand{\nn}{\nonumber\\}

\newcommand{\h}{\hspace}
\newcommand{\be}{\begin{equation}}
\newcommand{\e}{\end{equation}}
\newcommand{\aln}[1]{\begin{align}#1\end{align}}
\newcommand{\rf}[1]{(\ref{#1})}

\begin{document}
\title{
\vbox{
\baselineskip 14pt
\hfill \hbox{\normalsize KUNS-2647
}} \vskip 1cm
\bf \Large  A note on graviton exchange in emergent gravity scenario
\vskip 0.5cm
}
\author{
Hikaru~Kawai\thanks{E-mail: \tt hkawai@gauge.scphys.kyoto-u.ac.jp},~
Kiyoharu~Kawana\thanks{E-mail: \tt kiyokawa@gauge.scphys.kyoto-u.ac.jp}, and
Katsuta~Sakai\thanks{E-mail: \tt katsutas@gauge.scphys.kyoto-u.ac.jp} 
\bigskip\\
\it \normalsize
 Department of Physics, Kyoto University, Kyoto 606-8502, Japan\\
\smallskip
}
\date{}

\setcounter{page}{0}
\maketitle\thispagestyle{empty}
\abstract{\normalsize
It is well-known that there exists a close relation between a large $N$ matrix model and noncommutative (NC) field theory: The latter can be naturally obtained from the former by expanding it around a specific background. 
Because the matrix model can be a constructive formulation of string theory, 
this relation suggests that NC field theory can also include quantum gravity. 
In particular, the NC $U(1)$ gauge theory attracts much attention because 
its low-energy effective action partially contains 
gravity 
where the metric is determined by the $U(1)$ gauge field. 
Thus, the NC $U(1)$ gauge theory could be a quantum theory of gravity. 
In this paper, we investigate the scenario by calculating the scattering 
amplitude of massless test particles, and find that the NC $U(1)$ gauge theory correctly reproduces the amplitude of the usual graviton exchange if the noncommutativity that corresponds to the background of the matrix model is appropriately averaged. Although this result partially supports the relation between the NC $U(1)$ gauge theory and 
gravity, 
it is desirable to find a mechanism by which such an average is naturally realized.
}

\section{Introduction}
\label{introduction}
\quad Although superstring theory is one of the candidates that include quantum gravity,  
its perturbative formulation seems to be lack of predictability in the sense that there is no criteria about 
the favored choice of the background or the vacuum. Therefore, we need its non-perturbative formulation or a new framework that includes gravity to overcome this situation.
Among the various candidates, the IIB matrix model \cite{Ishibashi:1996xs,Aoki:1998bq} is promising because it can be 
the constructive formulation of type IIB string theory.
In the theory, the Wilson loops are identified with asymptotic string states 
in the continuum limit \cite{Fukuma:1997en,Hamada:1997dt}, 
and their Schwinger-Dyson equation (loop equation) represents 
the time evolution of strings, including their splitting and combination. 
Furthermore, in addition to the original IIB matrix model, many matrix models have been investigated such as 
its bosonic part alone and its deformation by adding new terms or matters.
In spite of these progresses, the meaning of matrix is not yet clear, and many interpretations have been proposed so far \cite{Aoki:1998vn,Iso:1999xs,Imai:2003vr,Imai:2003jb,Kaneko:2005pw,Kim:2011cr,Asano:2012mn,Kawai:2013wwa,Hanada:2005vr,Kawai:2007zz,Kawana:2016tkw,Hamada:2015dja}.
In this paper, we interpret the matrices as momentum because it is naturally related to field theory with the flat metric $g^{\mu\nu}=\delta^{\mu\nu}$ when we expand it around a specific background. See section \ref{U(1)gravity} for the details.

In particular, it is well known that there is a close relation between the matrix model and 
noncommutative (NC) field theory \cite{Aoki:1999vr,Ishibashi:1999hs,Iso:2000ew}: 
The fluctuation of matrices around 
some noncommutative classical solution is equivalent to the field of NC field theory, 
where the product of fields is defined by the star product. 
Remarkably, the NC $U(1)$ gauge field is uniformly coupled with all the matters. 
Such a special behavior of the $U(1)$ gauge field reminds us the property of gravity \cite{Rivelles:2002ez,Yang:2004vd,Yang:2006hj}. 
In fact, the coupling between the $U(1)$ gauge field and the matters 
can be expressed by the effective metric in the leading order with respect to 
the noncommutativity (see section \ref{U(1)gravity} for the details). 
Thus, the fluctuation of the $U(1)$ gauge field can be viewed as the fluctuation 
of the metric. Moreover, it is suggested that UV/IR mixing emerging 
at one-loop calculation of the $U(1)$ NC field can be understood in terms of induced or emergent gravity \cite{Steinacker:2007dq,Steinacker:2008ri,Grosse:2008xr,Kuntner:2011rr,Steinacker:2012ct}\!
\footnote{
Recently, another mechanism of emergent gravity was also discussed on a specific background \cite{Steinacker:2016vgf}.
}\!
(for a review, see also
\cite{Steinacker:2010rh,Steinacker:2011ix}).

While such successful results are present, it is not yet clear whether the 
mechanism rigorously reproduces the real gravity. For example, 
we have not yet obtained the explicit diffeomorphism invariance within the degrees of freedom of matrices.\footnote{On the other hand, the diffeomorphism can be explicitly seen in other interpretations of matrix model such as the covariant derivative interpretation \cite{Hanada:2005vr,Kawai:2007zz}.
} 
At a first glance, it seemingly does not work because of the off-shell degrees of freedom; the $U(1)$ gauge field has only four, while a graviton does ten. 
However, still there is the possibility because the off-shell degrees of freedom are unphysical.\footnote{
There is discussion about quantum gravity 
in terms of the $U(1)$ gauge field. See for example
\cite{Yang:2013aia}. 
} 
Therefore, it is quite necessary to check whether the emergent gravity scenario can actually explain the results of the ordinary gravity.
As a fist step, it is interesting and meaningful to calculate a two-body scattering amplitude 
of test particles exchanging the NC $U(1)$ gauge field, and 
to compare it with that of the usual graviton exchange. In this paper, we perform such analysis, 
and see that the NC $U(1)$ gauge theory correctly reproduces the usual graviton exchange if the noncommutativity is appropriately averaged 
and the test particles are massless (See Section \ref{sec:scattering} for the details.). 
Although this result shows a partial success of the mechanism, it may also indicate 
the necessity of considering another framework in order to produce the correct 
four-dimensional gravity. If such a new framework is actually found, 
this scenario becomes more and more promising, and we can get deep understating between 
the matrix model and gravity. 

In the following discussion, we assume that the flat metric is Euclidean.

\section{Brief review of emergent gravity}
\label{U(1)gravity}
\quad Before going into the calculation of the scattering amplitude, let us briefly review the emergent gravity scenario starting from the matrix model. 
We consider the following action: 
\aln{
 S&=S_{\text{IIB}}^{}+S_{\Phi}\nonumber
 \\
  &=-\mr{Tr}\paren{\frac{(2\pi)^2}{4\Lambda^4}\delta^{ac}\delta^{bd}
     [P_a^{},P_b^{}][P_c^{},P_d^{}]} -\frac{(2\pi)^2}{g^2\Lambda^4}\mr{Tr}\paren{\frac{1}{2}\delta^{ab}[P_a^{},\Phi]
     [P_b^{},\Phi]}, ~~(a,b=1,\cdots,10)
\label{action0}
}
where $S_{\text{IIB}}^{}$ is the bosonic part of the IIB matrix model, $\Lambda$ represents a cut-off scale, and $P_{a}^{}$ and $\Phi$ are $N\times N$ hermitian matrices. 
Note that $\delta^{ab}$ stands for the ten-dimensional flat metric. 
In the following discussion, we consider fluctuations around a specific background
$\bar{P}_a$. We interpret $\bar{P}_a$ as the derivatives with respect to 
the coordinate. 
This is called the momentum interpretation of the matrix model. 
In this model, $\bar{P}_{a}^{}$'s are determined by the classical equation of motion:
\be \left[P^b,\left[P_b,P_a^{}\right]\right]+\left[\Phi,\left[P_a^{},\Phi\right]\right]=0.\label{eq:eom}
\e
Among the various solutions, 
we consider the following one that gives the four dimensional NC spacetime: 
\aln{
 &[\bar{P}_\mu^{},\bar{P}_\nu^{}] = i\Lambda_{\text{NC}}^{2}\times \tilde{\theta}_{\mu\nu}{\bf 1}~~(\mu,\nu=1,\cdots,4), \nn
 &\bar{P}_i^{} = 0~~(i=5,\cdots,10),\ \Phi=0
\label{bg}
}
where $\tilde{\theta}_{\mu\nu}^{}$ is an antisymmetric dimensionless constant, 
and $\Lambda_{\text{NC}}$ is a NC scale. 
This is the well-known noncommutative geometry called the 4D Moyal-Weyl plane 
$\mathbb{R}^4_{\theta}$. 
By considering the fluctuation $A_\mu^{}:=P_\mu^{}-\bar{P}_\mu^{}$ around the solution, and using the well-known correspondence between matrix and function on the NC spacetime \cite{Aoki:1999vr}, the action takes the form 
\aln{
 S &= -\mr{Tr}\paren{\frac{(2\pi)^2}{4\Lambda^4}\delta^{\mu\alpha}\delta^{\nu\beta}
     [\bar{P}_\mu^{}+A_\mu^{},\bar{P}_\nu^{}+A_\nu^{}][\bar{P}_\alpha^{}+A_\alpha^{},\bar{P}_\beta^{}+A_\beta^{}]}
 -\frac{(2\pi)^2}{g^2\Lambda^4}\mr{Tr}\paren{\frac{1}{2}\delta^{\mu\nu}
     [\bar{P}_\mu^{}+A_\mu^{},\Phi][\bar{P}_\nu^{}+A_\nu^{},\Phi]}\nonumber
     \\
     &=\frac{\Lambda_{\text{NC}}^4}{\Lambda^4}\Bigg\{\frac{1}{4}N\tilde{\theta}^{\mu\nu}\tilde{\theta}_{\mu\nu}^{}
     + \frac{1}{4} \int d^4x\,\sqrt{|\tilde{\theta}|}\, 
    \left( \delta^{\mu\nu}\delta^{\alpha\beta} F_{\mu\alpha}^{}F_{\nu\beta}^{}\right)_\star^{}\nonumber
    \\
    &\h{6cm}+\frac{1}{2g^2}\int d^4x\,\sqrt{|\tilde{\theta}|}
     \delta^{\mu\nu}\Bigl(\paren{\partial_\mu\Phi-i[A_\mu,\Phi]}
     \paren{\partial_\nu\Phi-i[A_\nu,\Phi]}\Bigr)_\star^{}\Bigg\},
\label{actionNC}
}
where we have neglected the fluctuation of  $P_i^{}$'s ($i=5,\cdots10,$) for simplicity. Here, $\tilde{\theta}=\text{det}(\tilde{\theta}_{\mu\nu}^{})$, $F_{\mu\nu}=\partial_\mu^{}A_\nu^{}-\partial_\nu^{}A_\mu^{}-i\left[A_\mu^{},A_\nu^{}\right]_\star^{}$, and $\star$ represents 
the Moyal product defined by 
\aln{
 (f\star g)(x) 
     = \exp\left(\frac{i}{2}\Lambda_{\text{NC}}^{-2}\theta^{\mu\nu}\partial^{(y)}_\mu\partial^{(z)}_\nu\right)f(y)g(z)\bigg|_{y=z=x}^{}, 
}
where $\theta^{\mu\nu}=(\tilde{\theta}^{-1})^{\mu\nu}$.
Furthermore, note that we can always eliminate 
$(\Lambda_\text{NC}^4/\Lambda^4)|\tilde{\theta}|^{1/2}$ 
by the field redefinition 
$\Phi\rightarrow (\Lambda^2/\Lambda_\text{NC}^2)|\tilde{\theta}|^{-1/4}\Phi$ 
in the last term of Eq.(\ref{actionNC}). 
The above argument is the usual interpretation of the matrix model as NC field theory. 
On the other hand, it was also argued that $A_\mu^{}$ can be interpreted as the 
fluctuation of the four dimensional spacetime metric in the semi-classical limit 
\cite{Rivelles:2002ez,Yang:2004vd,Yang:2006hj}. Here, `semi-classical' means that we should keep the lowest order terms in $\Lambda_{\text{NC}}^{-2}\theta^{\mu\nu}$, and neglect higher order terms. 
In this approximation, noncommutativity gets switched off, 
and commutator 
turns into the Poisson bracket as 
\aln{
 [f,g]_\star \sim i\{f,g\}, ~~\{f,g\} \equiv \Lambda_{\text{NC}}^{-2}\times\theta^{\mu\nu}\partial_\mu f\partial_\nu g.
}
Then, Eq.(\ref{actionNC}) now becomes
\aln{ S \bigg|_{\text{semi}}= 
        &
        \frac{\Lambda_\text{NC}^4}{4\Lambda^4} \int d^4x\,\sqrt{|\tilde{\theta}|}\, 
   (\sqrt{|G|}G^{\mu\nu})\tilde{\theta}_{\mu\alpha}^{}\tilde{\theta}_{\nu\beta}^{}(\sqrt{|G|}G^{\alpha\beta})
  +\int d^4x\,\frac{1}{2g^2}\sqrt{|G|}G^{\mu\nu}\partial_\mu^{}\Phi\partial_\nu^{}\Phi, 
\label{eq:semiaction}
}
where
\aln{
\sqrt{|G|}G^{\mu\nu}=\delta^{\mu\nu}+&
\Lambda_{\text{NC}}^{-2}\times\left(\theta^{\alpha\mu}\partial_\alpha^{}A^\nu
+\theta^{\alpha\nu}\partial_\alpha^{}A^\mu\right) 
+
\Lambda_{\text{NC}}^{-4}\times {\cal{O}}(A^2), 
\label{eq:A to Phi}
}
From this we can read the fluctuation of the metric 
$h^{\mu\nu}=-(G^{\mu\nu}-\delta^{\mu\nu})$ as 
\aln{
h^{\mu\nu}=&\Lambda_{\text{NC}}^{-2}\times\left(
\theta^{\mu\alpha}
\partial_\alpha^{} A^{\nu}+\theta^{\nu\alpha}
\partial_\alpha^{} A^{\mu}
+\frac{1}{2}\delta^{\mu\nu}\theta^{\alpha\beta}F_{\alpha\beta}\right)+\Lambda_{\text{NC}}^{-4}\times {\cal{O}}(A^2)
\label{eq:naive metric perturbation}
}
In the following analysis, we investigate the dynamics of $A_\mu$ 
which is quadratic in the effective action. In this sense, the ${\cal{O}}(A^2)$ terms 
in $h^{\mu\nu}$ are not necessarily as long as we expand the Einstein-Hilbert (EH) action at the linearized level because such terms give higher order contributions.\footnote{
On the other hand, the ${\cal{O}}(A^2)$ terms 
in $h^{\mu\nu}$ are necessarily when we expand the cosmological constant term.
} 
From the above equations, one can actually see that $\Phi$ couples to $A_\mu^{}$ in the covariant way, and that $A_\mu^{}$ can be interpreted as the fluctuation of the metric. 
On the other hand, as for the bosonic part of IIB matrix model, its semi-classical action cannot be written in a covariant way. 
(See the first term in Eq.\rf{eq:semiaction}.) Although this is a big problem in the present formulation of emergent gravity, 
{we simply drop the term in the following discussion. 
In other words, we will focus on the matrix model which does not contain 
$F_{\mu\nu}^2\sim[P_\mu,P_\nu]^2$ term. Then as we will see below\rf{effaction}, such a term 
is not induced by the quantum correction due to the noncommutativity.
\\

Although we have found that the $U(1)$ gauge field can be understood as the fluctuation of the metric, we have not yet obtained the action for it. 
It was claimed that it is given by the induced EH action by considering the one-loop effective action 
of $A_\mu^{}$ in the semi-classical limit \cite{Grosse:2008xr}.\footnote{Here, note that there exist two ways of obtaining such a semi-classical limit: One is to take the semi-classical limit after calculating the one-loop effective action as a NC field theory. The other is to take first the semi-classical limit at the tree-level action, and calculate the effective action as an ordinary field theory. However, we have checked that both of the approaches produce the same result.
} 
By calculating the scalar one-loop diagrams (Fig.\ref{fig:one-loop}), we obtain the effective action of $A_\mu^{}$ as a NC field theory \cite{Grosse:2008xr}:\footnote{
In this reference, calculation was done by adding the mass term for the scalar as a regulator, 
and then taking the massless limit. Furthermore, the following replacement is used as a regularization of the loop integral:
\be \int \frac{d^4 p}{(2\pi)^4}\frac{f(p)}{[p^2+\bigtriangleup^2]^2}\rightarrow\int_0^{\infty}d\alpha\alpha\int \frac{d^4 p}{(2\pi)^4}f(p) e^{-\alpha(p^2+\bigtriangleup^2)-1/(\alpha\Lambda^2)}.
\e
Therefore, to maintain the consistency, we also use this regularization scheme in the following calculation.
} 
\begin{figure}
\begin{center}
\includegraphics[width=14cm,clip]{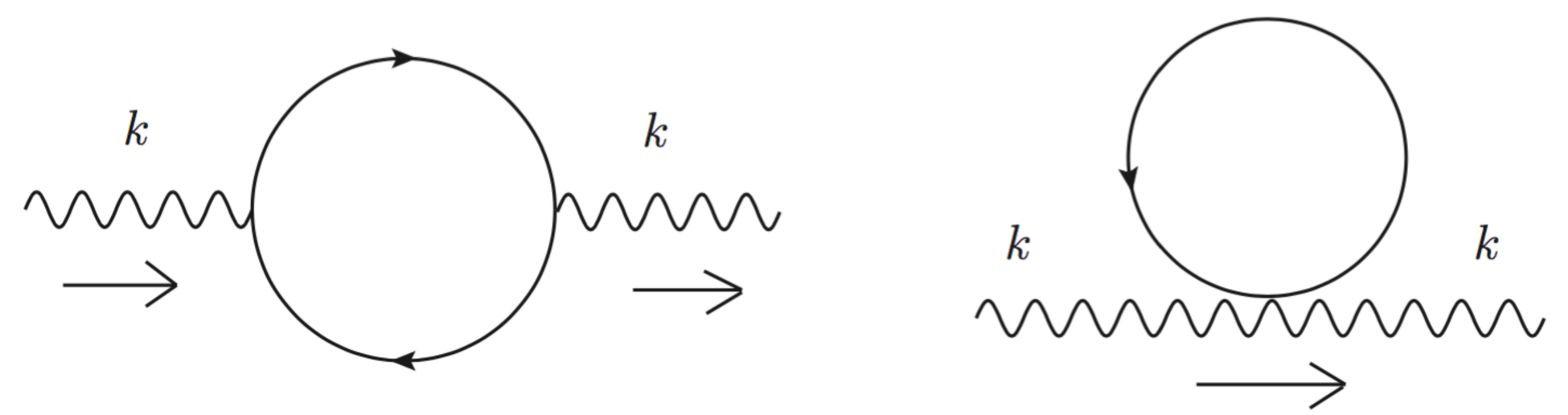}
\caption{One-loop diagrams needed for the computation of the effective action of the NC $U(1)$ gauge field. }
\label{fig:one-loop}
\end{center}
\end{figure}
%
\aln{
 e^{-\Gamma_\Phi} =& \int \mathcal{D}\Phi e^{-S}\big|_\text{1-loop without IIB action},\\
 \Gamma_\Phi =& -\frac{1}{32\pi^2g^2}\int\frac{d^4p}{(2\pi)^4}
     \biggl[-\frac{1}{6}F_{\mu\nu}(p)F^{\mu\nu}(-p)
     \log\paren{\frac{\Lambda^2}{\Lambda_\mr{eff}^2}}\nn
     &~~~~+\frac{1}{4}\theta^{\mu\nu}F_{\mu\nu}(p)\theta^{\lambda\rho}F_{\lambda\rho}(-p)
     \paren{\Lambda_\mr{eff}^4-\frac{1}{6}p^2\Lambda_\mr{eff}^2+\frac{(p^2)^2}{1800}
     \paren{47-30\log\paren{\frac{p^2}{\Lambda_\mr{eff}^2}}}}\biggr],
\label{NCeffectiveaction}
}
where $\Lambda_\mr{eff}^{-2}=\Lambda^{-2}+\tilde{p}^2/(4\Lambda_\mr{NC}^4)$, 
$\tilde{p}^\mu=\theta^{\mu\nu}p_\nu$,  
and $\Lambda$ is the cutoff momentum for loop integral. 
We suppose $\tilde{p}^2$ and $p^2$ are the same scale,  
since $\theta^{\mu\nu}$ is dimentionless and expected to be $O(1)$. 
When we focus on the IR regime, 
\aln{
 \frac{p^2\Lambda^2}{\Lambda_\mr{NC}^4} < 1, 
\label{IR}
}
Eq.\rf{NCeffectiveaction} can be expanded as
\aln{
 \Gamma_\Phi \sim&-\frac{1}{32\pi^2g^2}\int\frac{d^4p}{(2\pi^4)}\biggl[
     \frac{\Lambda^4}{4\Lambda_\text{NC}^4}
     \theta^{\mu\nu}F_{\mu\nu}(p)\theta^{\lambda\rho}F_{\lambda\rho}(-p)
     -\frac{\Lambda^4}{\Lambda_\text{NC}^4}\frac{\Lambda^2}{8\Lambda_\text{NC}^4}
     \tilde{p}^2\theta^{\mu\nu}F_{\mu\nu}(p)\theta^{\lambda\rho}F_{\lambda\rho}(-p)\nn
 &~~~~~~~~~~~~~~~~~~~~~~~~~~~-\frac{\Lambda^2}{24\Lambda_\mr{NC}^4}
     \left(F^{\mu\nu}(p)F_{\mu\nu}(-p)\tilde{p}^2
     +\theta^{\mu\nu}F_{\mu\nu}(p)
     \theta^{\lambda\rho}F_{\lambda\rho}(-p)p^2\right)\biggr]\nn
 =&-\frac{1}{32\pi^2g^2}\int d^4x\biggl[\frac{\Lambda^4}{4\Lambda_\text{NC}^4}
     \theta^{\mu\nu}F_{\mu\nu}\theta^{\lambda\rho}F_{\lambda\rho}
    +\frac{\Lambda^4}{\Lambda_\text{NC}^4}\frac{\Lambda^2}{8\Lambda_\text{NC}^4}
     \theta^{\mu\nu}F_{\mu\nu}(\partial\circ\partial)
     \theta^{\lambda\rho} F_{\lambda\rho}\nn
     &~~~~~~~~~~~~~~~~~~~~~~~~
     \h{3cm}+\frac{\Lambda^2}{24 \Lambda_\mr{NC}^4}\biggl(
     F^{\mu\nu}(\partial\circ\partial) F_{\mu\nu}
     +\theta^{\mu\nu}F_{\mu\nu}\Box\theta^{\lambda\rho} F_{\lambda\rho}\biggr)\biggr],
\label{effaction}
}
where $\partial\circ\partial=\theta^{\mu\alpha}\theta^{\nu\beta}\delta_{\alpha\beta}\partial_\mu\partial_\nu$, and $\Box=\delta^{\mu\nu}\partial_\mu^{}\partial_\nu^{}$. This result should be compared with the EH action with the cosmological constant term where the metric is given by Eq.(\ref{eq:naive metric perturbation}): 
\aln{ S_{G}&=
\frac{1}{16\pi^2}
\int d^4x \sqrt{|G|} \left( -\frac{1}{2}\Lambda^4
     -\frac{\Lambda^2}{12}R[G]\right)\nonumber
\\
&\sim-\frac{1}{32\pi^2g^2}\int d^4x\biggl[\frac{\Lambda^4}{4\Lambda_\text{NC}^4}
     \theta^{\mu\nu}F_{\mu\nu}\theta^{\lambda\rho}F_{\lambda\rho}
     +\frac{\Lambda^2}{24 \Lambda_\mr{NC}^4}
     F^{\mu\nu}\partial\circ\partial F_{\mu\nu}^{}\biggr]
     ,
     \label{eq:emergent action}
}
where we have extracted the quadratic part in $A_\mu$ and rescaled it as 
$A_\mu \rightarrow A_\mu/g$. From Eq.\rf{eq:emergent action}, one can see that 
the terms $\theta^{\mu\nu}F_{\mu\nu}\Box\theta^{\lambda\rho} F_{\lambda\rho}$ 
and $\theta^{\mu\nu}F_{\mu\nu}(\partial\circ\partial)
\theta^{\lambda\rho} F_{\lambda\rho}$ in Eq.\rf{effaction} 
are absent in Eq.\rf{eq:emergent action}. The latter is, however, 
a higher-order term in $(\Lambda/\Lambda_\text{NC})^4$. 
Here we consider the effects of the noncommutativity in its lowest order. 
In other words, we assume that $\Lambda_\text{NC}$ is larger than the cut-off momentum $\Lambda$ and we shall neglect this term in the following discussion.\footnote{
The following analysis works in parallel 
and gives qualitatively the same result 
even if we take this term into account.  
} 
The above mismatch between Eq.(\ref{effaction}) and Eq.(\ref{eq:emergent action})  originates in the path-integral measure: In the NC theory, it is induced from the flat metric in the functional space of $\Phi$. 
\be ||\delta \Phi||^2=\int d^4x\ \delta\Phi(x)^2,\label{eq:measure}
\e
and this apparently violates the diffeomorphism invariance. If we use the diffeomorphism transformation 
\aln{
x^\mu \rightarrow y^\mu = x^\mu-\theta^{\mu\nu}A_\nu, 
\label{changeofvariables}
}
which is not realized in the NC $U(1)$ gauge theory, 
we can make $h^{\mu\nu}$ traceless in the leading-order in $A_\mu$. In such coordinates, the one-loop effective action indeed matches the EH action, as discussed in \cite{Grosse:2008xr}. 
See \ref{app:diff} for the details.

In spite of the mismatch, the similarity between Eq.(\ref{effaction}) and Eq.(\ref{eq:emergent action}) is impressive, and it is meaningful to study whether the NC $U(1)$ gauge theory can actually describe the real gravity. 
In the following, we in particular consider the amplitude of the graviton exchange 
between two scalars.

\section{Does Noncommutative $U(1)$ gauge field actually describe gravity ?}\label{sec:scattering}
\quad As a first step, we compute the two-body scattering amplitude of the scalar particles exchanging the $U(1)$ gauge field whose action is given by Eq.(\ref{effaction}).
In the following discussion, we put $32\pi^2 g^2=1$ for simplicity, and drop the first term in Eq.(\ref{effaction}) 
because it corresponds to the 
cosmological constant (CC) term, 
which we assume to be canceled by some mechanism. Adding a gauge fixing term 
and rewriting them in terms of $A_\mu$, we have 
\aln{
 \Gamma_\Phi\big|_{O(\Lambda^2) \text{ without CC term}}^{} 
 &+\frac{1}{\alpha}\int\frac{d^4p}{(2\pi)^4}
 \frac{\Lambda^2}{12\Lambda_\text{NC}^4}\tilde{p}^2p^\mu A_\mu(p)p^\nu A_\nu(-p) 
\nn
 &= 
\frac{\Lambda^2 }{12 \Lambda_\mr{NC}^4}\int\frac{d^4p}{(2\pi)^4}A^\mu(p)\left[
 \tilde{p}^2\Bigl(p^2\delta_{\mu\nu}^{}
 -\left(1-\frac{1}{\alpha}\right)p^\mu p^\nu \Bigr)
+2p^2\tilde{p}^\mu\tilde{p}^\nu \right]
 A^\nu(-p), 
}
where $\alpha$ represents the gauge freedom. From this, one can read off the propagator of $A_\mu$ as 
\aln{
 D_{\mu\nu}(p) =
 \frac{6\Lambda_\mr{NC}^4}{\Lambda^2}\frac{1}{\tilde{p}^2p^2}\biggl[
 \delta_{\mu\nu}-(1-\alpha)\frac{p_\mu p_\nu}{p^2}
 -\frac{2}{3}\frac{\tilde{p}_\mu\tilde{p}_\nu}{\tilde{p}^2}
 \biggr].
\label{propagator}
}
Supposing $A_\mu$ to propagate with Eq.\rf{propagator}
, the two-body scattering amplitude of the scalar 
can be calculated in the semi-classical approximation. In the following discussion, we take the Feynman gauge 
$\alpha=1$. From Eqs.\rf{eq:semiaction} and \rf{eq:A to Phi} we can read 
the interaction between $\Phi$ and $A_\mu^{}$ as
\aln{
{\cal{L}}_{\text{Int}}=\Lambda_{\text{NC}}^{-2}\times 
\theta^{\alpha\mu}\partial_\alpha A^\nu\partial_\mu^{}\Phi\partial_\nu^{}\Phi
}
from which we can read the vertex as

\begin{minipage}{3cm}
\includegraphics[width=4cm]{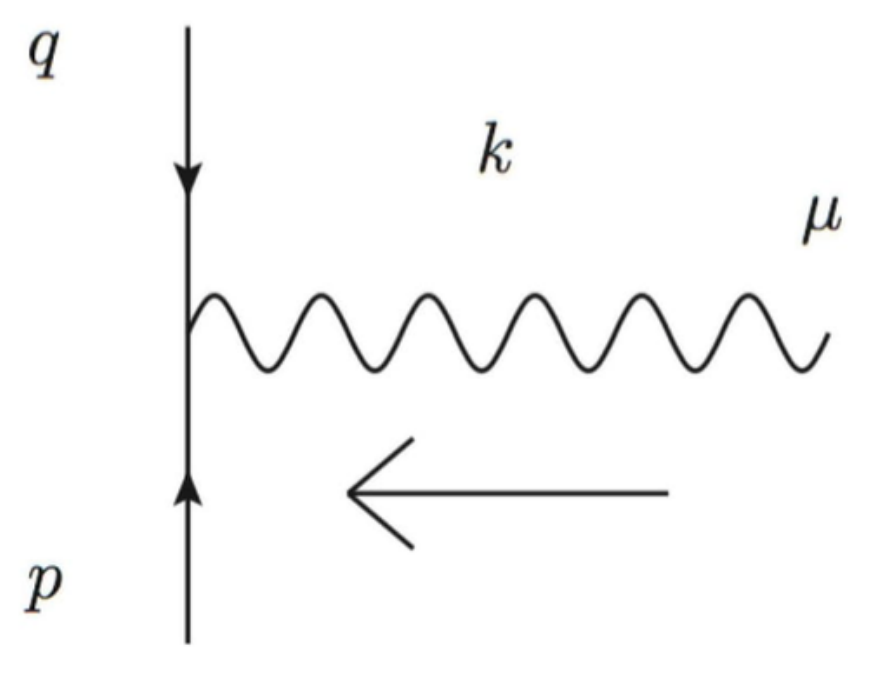}
\end{minipage}
\begin{minipage}{10cm}
\vspace{-0.5cm}
\aln{
=
i\Lambda_\text{NC}^{-2}\times\left[
p^\mu(q\cdot\tilde{k})+q^\mu(p\cdot\tilde{k})\right].
}
\end{minipage}
\\
\\
%
We can now compute the two-body scattering amplitude (see Fig.\ref{amplitude}). 
%
\begin{figure}
\begin{center}
\includegraphics[width=14cm]{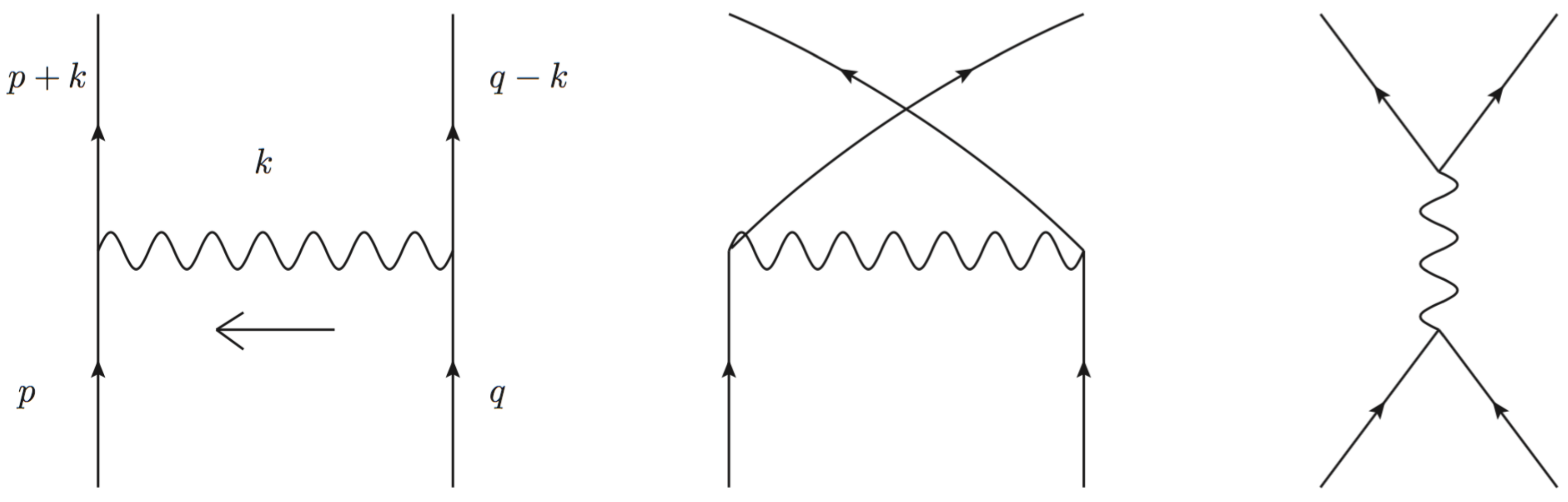}
\caption{A scattering of test particles exchanging the $U(1)$ filed or graviton. In the former case, we read its propagator from the one-loop effective action Eq.(\ref{effaction}).
}
\label{amplitude}
\end{center}
\end{figure}
%
%
Along with on-shell conditions for the scalar 
\aln{
 p^2=q^2=0,~~
 (p+k)^2 = p^2,~~(q-k)^2=q^2, 
\label{on-shell}
}
we obtain 
\aln{
 {\cal{M}}_A
  = \frac{6}{\Lambda^2}
 \frac{1}{k^2}\frac{(p\cdot\tilde{k})(q\cdot\tilde{k})}{\tilde{k}^2}\biggl[
 (4p\cdot q+k^2)
 -\frac{8}{3}\frac{(p\cdot\tilde{k})(q\cdot\tilde{k})}{\tilde{k}^2}\biggr]+(s\text{ channel})+(u\text{ channel}). 
\label{A-ampx}
}

This should be compared with the scattering amplitude calculated from the ordinary gravity system:
\aln{
 S
  =& S_{G}+S_\mr{gf}+S_{\Phi},\\
 S_{G} =&\frac{1}{2G_\mr{N}}\int d^4x\sqrt{|(\delta+h)|}
     R[\delta+h]\Big|_{\mr{quadratic\,part\,in\,}h},\nn
 =&\int d^4x\biggl[\frac{1}{8}\partial_\lambda h_{\mu\nu}\partial^\lambda h^{\mu\nu}
     -\frac{1}{4}\partial^\mu h_{\mu\nu}\partial_\lambda h^{\lambda\nu}
     +\frac{1}{4}\partial^\mu h_{\mu\nu}\partial^\nu h
     -\frac{1}{8}\partial_\mu h\partial^\mu h \biggr],\\
 S_\mr{gf} =&\frac{1}{4}\int d^4x\paren{\partial^\nu h_{\mu\nu}-\frac{1}{2}\partial_\mu h}
     \paren{\partial^\lambda h^\mu_{~~\lambda}-\frac{1}{2}\partial^\mu h},\\
 S_{\Phi} =&\int d^4x\sqrt{|(\delta+h)|}\biggl[\frac{1}{2}(\delta^{\mu\nu}
     -h^{\mu\nu})\partial_\mu\Phi\partial_\nu\Phi\biggr]
     \bigg|_{\mr{0th\,and\,1st\,order\,of\,}h}\nn
 =&\int d^4x\biggl[\frac{1}{2}\partial_\mu\Phi\partial^\mu\Phi
     -\frac{1}{2}h^{\mu\nu}\partial_\mu\Phi\partial_\nu\Phi
    +\frac{1}{4}h\partial_\mu\Phi\partial^\mu\Phi\biggr],
}
where the graviton is gauge-fixed in the de Donder gauge (harmonic gauge) which 
leads to the following propagator of graviton:
\aln{
 D^{(h)}_{\mu\nu\lambda\rho}(k) =\frac{2}{k^2}(\delta_{\mu\lambda}\delta_{\nu\rho}
     +\delta_{\mu\rho}\delta_{\nu\lambda}-\delta_{\mu\nu}\delta_{\lambda\rho}).
}
From the straightforward calculation, we obtain
\aln{
 {\cal{M}}_G^{} 
 &= 2G_\mr{N}\paren{\frac{2(p\cdot q)^2}{k^2}+p\cdot q}+(s\leftrightarrow t)+(t\leftrightarrow u)=-G_N^{}\left(\frac{su}{t}+\frac{tu}{s}+\frac{ts}{u}\right), 
\label{g-amp}
}
where $s$, $t$ and $u$ are the Mandelstam variables. This does not match Eq.\rf{A-ampx} 
although both of them lead to the inverse-square law. 
In particular, $\theta^{\mu\nu}$ explicitly remains in Eq.\rf{A-ampx}, 
and we need some mechanism to eliminate it. 
Because $\theta^{\mu\nu}$ is a moduli parameter which specifies the classical solution, it 
is natural to take a kind of average over it. 
For example, let us consider the average over the direction of $\theta^{\mu\nu}$ with 
the `absolute value of $\theta^{\mu\nu}$' being fixed: 
\be \theta^{\mu\alpha}\theta^{\nu\beta}\delta_{\alpha\beta} 
= \delta^{\mu\nu}. \label{theta to delta}\e
We assume that $\theta^{\mu\nu}$ distributes in the Lorentz covariant manner: 
\be \theta^{\mu\nu} \rightarrow \theta_M^{\mu\nu} = 
M^\mu_{~\alpha} M^\nu_{~\beta} \theta^{\alpha\beta} \label{thetaLorentz}, \e
where $M^\mu_{~\alpha}$ is an element of $SO(4)$. 
This is compatible with the assumption \rf{theta to delta}. 
Then the average over the direction of $\theta^{\mu\nu}$ yields Lorentz covariant 
quantities: 
\aln{
\int_{SO(4)}dM~\theta_M^{\mu\nu}\theta_M^{\lambda\rho} &= 
\frac{1}{3}(\delta^{\mu\lambda}\delta^{\nu\rho}-\delta^{\mu\rho}\delta^{\nu\lambda})
\equiv\frac{1}{3}\Delta^{\mu\nu\lambda\rho}, \nn
\int_{SO(4)}dM~\theta_M^{\mu\nu}\theta_M^{\lambda\rho}
\theta_M^{\alpha\beta}\theta_M^{\gamma\delta}
&=-\frac{1}{27}(
\Delta^{\mu\nu\lambda\rho}\Delta^{\alpha\beta\gamma\delta}
+\Delta^{\mu\nu\alpha\beta}\Delta^{\lambda\rho\gamma\delta}
+\Delta^{\mu\nu\gamma\delta}\Delta^{\lambda\rho\alpha\beta}) \nn
&+\frac{1}{9}\Bigl\{
(\delta^{\nu\lambda}\delta^{\rho\alpha}\delta^{\beta\gamma}\delta^{\delta\mu}
+\delta^{\nu\alpha}\delta^{\beta\gamma}\delta^{\delta\lambda}\delta^{\rho\mu}
+\delta^{\nu\gamma}\delta^{\delta\lambda}\delta^{\rho\alpha}\delta^{\beta\mu}) 
+[\mu\nu][\lambda\rho][\alpha\beta][\gamma\delta]\Bigr\}, 
\label{thetareplace}
}
where $dM$ denotes the Haar measure of $SO(4)$, and 
$[\mu\nu][\lambda\rho][\alpha\beta][\gamma\delta]$
represents the antisymmetrized terms of the first one with respect to the superscripts 
in each of the brackets. Here the coefficients in the RHS of Eq.\rf{thetareplace} are 
determined so that they are consistent with Eq.\rf{theta to delta}. 
After taking such an average,} Eq.(\ref{A-ampx}) now becomes
\aln{
{\cal{M}}_A 
=\frac{2}{\Lambda^2}\biggl[
\frac{52}{27}\frac{1}{k^2}2(p\cdot q)^2+\frac{14}{27}p\cdot q+ \frac{1}{36}k^2\biggr]+(s\leftrightarrow t)+(t\leftrightarrow u)
=-\frac{52}{27\Lambda^2}\left(
\frac{su}{t}+\frac{tu}{s}+\frac{st}{u}
\right), 
\label{A-ampxreduced}
}
which correctly reproduces Eq.\rf{g-amp}.

Therefore, if $\theta^{\mu\nu}$ is appropriately averaged as Eq.(\ref{thetareplace}), 
the scattering amplitude in the induced gravity scenario coincides 
with that of the ordinary gravity. 
The question is the meaning and validity of averaging $\theta^{\mu\nu}$. 
In the analysis above, we first calculated the amplitude with a fixed $\theta^{\mu\nu}$ 
and then averaged over its direction. 
On the other hand, turning back to the matrix model, 
$\theta^{\mu\nu}$ is determined by the commutator of matrices, and the path integral 
over them naturally includes the integration over $\theta^{\mu\nu}$ as a fundamental variable. 
In the language of NC field theory, this implies that $\theta^{\mu\nu}$ could 
have independent fluctuation, and that the average over $\theta^{\mu\nu}$ 
corresponds to the integral over 
$\theta^{\mu\nu}$: 
\aln{ Z&= \int d\theta f(\theta)\int{\cal{D}}A\int{\cal{D}}\Phi \exp\left(-S\right), 
\label{eq:ordinal interpretation}
}
where $f(\theta)$ is some weight function. 
In order to justify this picture, it is necessary to check whether the average 
as Eq.\rf{thetareplace} 
can actually produce the correct results in other scattering processes of gravity.

Further investigation is needed on the treatment of $\theta^{\mu\nu}$, with the 
emphasis on its degrees of freedom. For example, 
lifting $\theta^{\mu\nu}_{}$ as an independent field is attractive in the aspect of degrees of 
freedom. It is possible to compensate the discrepancy between the off-shell degrees of 
freedom of the 
NC $U(1)$ gauge field and the ordinary gravity, because $\theta^{\mu\nu}$ has 
six independent components. 
We can also consider a new matrix model or novel interpretation of matrix variables
in which gravity does not necessarily come from the noncommutativity\cite{Hanada:2005vr}\cite{Steinacker:2016vgf}.

\section{Summary}
\label{discusion}
\quad We have investigated the emergent gravity scenario by examining the two-body scattering of the scalar particles exchanging the NC $U(1)$ gauge field. 
As long as we take Eq.\rf{A-ampx} literally, 
the fundamental force acting between test particles looks somewhat different 
from that of the ordinary gravity, although the NC $U(1)$ gauge field can be 
viewed as the metric fluctuation. 
However, once we take the average over the direction of $\theta^{\mu\nu}$, 
the resulting amplitude matches that of the ordinary gravity.  
The origin of the averaging procedure can be attributed to the path integral 
of the matrices in the matrix model. 
It is interesting to investigate this possibility further.

\section*{Acknowledgement} 
K.K is supported by Japan Society for the Promotion of Science (JSPS) Fellowship for Young Scientists.

\appendix
 
\def\thesection{Appendix} 
\section{Change of effective action and amplitude by diffeomorphism}\label{app:diff}
One might take seriously the discrepancy between the two actions, Eqs. \rf{effaction} 
and \rf{eq:emergent action}. If some procedure makes the one-loop effective action 
identical to Eq.\rf{eq:emergent action}, then one can regard the 
propagation of the NC $U(1)$ gauge field as that of the metric fluctuation
at the action level. For example, let us consider the diffeomorphism 
transformation \rf{changeofvariables}. Then the metric fluctuation $h^{\mu\nu}$ transforms as
\aln{
\delta h^{\mu\nu}&=\partial^\mu\xi^\nu+\partial^\nu\xi^\mu, \\
\xi^\mu&=\theta^{\mu\alpha}A_\alpha^{}, 
}
although the diffeomorphism transformation is not actually realized in the 
NC $U(1)$ gauge theory. Assuming that the above change of $h^{\mu\nu}$ is verified, we obtain 
\aln{
 h^{\mu\nu}=&\Lambda_{\text{NC}}^{-2}\times\left(
\theta^{\mu\alpha}F_\alpha^{~\nu}
+\theta^{\nu\alpha}F_\alpha^{~\mu}
+\frac{1}{2}\delta^{\mu\nu}\theta^{\alpha\beta}F_{\alpha\beta}\right)+\Lambda_{\text{NC}}^{-4}\times {\cal{O}}(A^2). 
\label{h-y}
} 
In the viewpoint of the NC $U(1)$ gauge theory, such a transformation 
produces additional interaction terms to the action. If we take this new action as the tree-level action, 
one-loop effective action of $A_\mu$ is given by 
\aln{
\Gamma_{\Phi}^{(y)}
     &\sim-\frac{1}{32\pi^2g^2}\int d^4y\biggl[\frac{\Lambda^4}{4\Lambda_\text{NC}^4}
     \theta^{\mu\nu}F_{\mu\nu}\theta^{\lambda\rho}F_{\lambda\rho}
     +\frac{\Lambda^2}{24 \Lambda_\mr{NC}^4}
     F^{\mu\nu}\partial\circ\partial F_{\mu\nu}
     \biggr], \ 
\label{y-oneloop}
}
where we extract the quadratic parts in $A_\mu$ and drop higher-order terms in 
$\Lambda^2/\Lambda_\text{NC}^2$. This coincides with the expanded EH action 
Eq.\rf{eq:emergent action}. This can be understood as follows. In the coordinates yielding 
Eq.\rf{h-y}, the leading term in $A_\mu$ vanishes in the trace of $h^{\mu\nu}$. Therefore 
the diffeomorphism-invariant measure in the functional space approximately agree 
with the flat measure: 
\be ||\delta\Phi||^2=\int d^4x\ \sqrt{G}\delta\Phi(x)^2 \simeq 
\int d^4x\delta\Phi(x)^2, \e
As a result, the path integral over $\Phi$ in the NC $U(1)$ gauge theory gives 
the diffeomorphism-invariant effective action Eq.\rf{eq:emergent action}, as far as we keep track of 
the lowest-order in $A_\mu$.

Once we adopts the second term in Eq.\rf{y-oneloop} as the kinetic term of $A_\mu$, 
we can do similar calculation to that in section 3. By considering the 
interaction corresponding to Eq.\rf{h-y}, the scattering amplitude is given by 
\aln{
 {\cal{M}}_A^{(y)} = \frac{6}{\Lambda^2}
 \frac{1}{k^2\tilde{k}^2}
(p\cdot\tilde{k})(q\cdot\tilde{k})(4p\cdot q+k^2)+(s\text{ channel})+(u\text{ channel}). 
\label{A-ampy}
}
This result shows that the amplitude does not match Eq.\rf{g-amp} 
even if the effective action agrees with the action obtained from 
the EH action after the substitution \rf{h-y}. 
It is mainly because of the lack of the degrees of freedom. However, if we assume 
the average leading to the average \rf{thetareplace}, 
we can obtain the result identical to the gravitational one: 
\aln{{\cal{M}}_A^{(y)}
=\frac{2}{\Lambda^2}\biggl[
2\left\{\frac{1}{k^2}2(p\cdot q)^2+p\cdot q\right\}+\frac{1}{4}k^2\biggr]+(s\leftrightarrow t)+(t\leftrightarrow u)=-\frac{2}{\Lambda^2}\left(\frac{su}{t}+\frac{tu}{s}+\frac{ts}{u}\right).
\label{A-ampyreduced}
}



\end{document}